\documentstyle[prd,aps,preprint]{revtex}

\def\beq{\begin{equation}}
\def\eeq{\end{equation}}

\begin{document}
\tightenlines

\title{\hfill {\rm \normalsize hep-th/0005193} \medskip \\ 
On Brane-World Cosmology}

\author{Yu.~V.~Shtanov\footnote{E-mail: shtanov@ap3.bitp.kiev.ua}} 
\address{Bogolyubov Institute for Theoretical Physics, Kiev 03143, Ukraine}

\date{May 22, 2000}

\maketitle

\thispagestyle{empty}

\sloppy

\begin{abstract}
 In the study of three-brane cosmological models, an unusual law of 
cosmological expansion on the brane has been reported. According to this law,
the energy density of matter on the brane quadratically enters the 
right-hand side of the new equations for the brane world, in contrast with 
the standard cosmology, where it enters the similar equations linearly.  
However, this result is obtained in the absence of curvature-dependent terms 
in the action for the brane.  In this paper, we derive the field equations 
for a brane world embedded into a five-dimensional spacetime in the case where 
such terms are present.  We also discuss some cosmological solutions of the 
resulting equations.  
\end{abstract}


\section{Introduction}

 The string-theory inspired idea of a four-dimensional universe 
as a brane world embedded into a multi-dimensional universe has now become 
very popular.  In many papers, models of such kind are considered in the 
four-dimensional cosmological context; in particular, the theory of 
cosmological perturbations within the frames of such models is currently 
being developed (for a comprehensive list of references see, e.g., 
\cite{vdBDBL}).  In several papers \cite{LOW,BDL,BDEL,BV}, an unusual law 
of cosmological expansion of a four-dimensional universe embedded into 
(or bounding \cite{BV}) a five-dimensional space has been reported. According 
to this law, the energy density of matter on the brane enters quadratically 
the right-hand side of the new equations for the brane world, in contrast 
with the standard cosmology, where it enters the similar equations linearly.  
Such a behaviour might strongly modify the standard cosmological model 
\cite{BDEL}.  There exist several approaches to solving this problem. 
Thus, for example, in \cite{CGKT,CGS,BDEL} it was shown that the 
standard cosmological evolution can be recovered in models with a large
five-dimensional cosmological constant and brane tension and, in 
\cite{CGRT}, it was also pointed out that, in the presence of a mechanism
for stabilization of the radius of the extra dimension, the cosmologies are 
generically of the Friedmann--Robertson--Walker ordinary four-dimensional 
form.  

 However, the basic equations of the theory were thus far  
obtained in the absence of curvature-dependent terms in the action for the 
four-dimensional brane world.  The purpose of this article 
is to derive the field equations for a brane world embedded into 
(or bounding) a five-dimensional space in the case where such 
curvature-dependent terms are present.  After that, we consider a particular 
example of a cosmological situation and discuss some of its solutions.

\section{Field equations}

 Consider a theory with a four-dimensional hypersurface (brane) 
$\Sigma$ which is the boundary of a five-dimensional manifold 
${\cal M}$.\footnote{In fact, all the equations of this section are 
valid for a $n$-dimensional space 
${\cal M}$ bounded by a $(n\!-\!1)$-dimensional hypersurface $\Sigma$.} 
We take the action of the theory to have the natural form
\beq \label{action}
S = M_5^3  \left[ \int_{\cal M} \left({}^{(5)}R  
- 2 \Lambda_5 \right) + 2 \int_{\Sigma} K  \right] + \int_{\cal M} 
L_5 (g_{ab}, \Phi)  
+ M_4^2 \int_{\Sigma} \left({}^{(4)}R - 
2 \Lambda_4 \right) + \int_{\Sigma} L_4 (h_{ab}, \phi) \, . 
\eeq
Here, ${}^{(5)}R$ is the scalar curvature of the Lorentzian 
five-dimensional metric $g_{ab}$ on ${\cal M}$, and ${}^{(4)}R$ is the scalar 
curvature of the induced metric $h_{ab} = g_{ab} - n_a n_b$ on $\Sigma$, 
where $n^a$ is the vector field of the outer unit normal to $\Sigma$.
The boundary $\Sigma$ is assumed to be timelike, so that the vector field 
$n^a$ is spacelike. The quantity $K = K_{ab} h^{ab}$ is the trace of the 
symmetric tensor of extrinsic curvature 
$K_{ab} = h^c{}_a \nabla_c n_b$ of $\Sigma$ in ${\cal M}$. The symbols 
$L_5 (g_{ab}, \Phi)$ and $L_4 (h_{ab}, \phi)$ denote, respectively, the 
Lagrangian densities of the five-dimensional matter fields $\Phi$ and of the
four-dimensional matter fields $\phi$ whose dynamics is restricted to the 
boundary $\Sigma$ so that they interact only with the induced metric $h_{ab}$.
Note that some of the fields $\phi$ in principle may represent restrictions 
of some of the fields $\Phi$ to the boundary $\Sigma$. 
All integrations over ${\cal M}$ and over $\Sigma$ are taken, 
respectively, with the natural volume elements $\sqrt{- g}\, d^5 x$ and 
$\sqrt{- h}\, d^4 x$, where $g$ and $h$ are, respectively, the determinants 
of the matrices of components of the metric on ${\cal M}$ and of the 
induced metric on $\Sigma$ in a coordinate basis.  The symbols $M_n$ and 
$\Lambda_n$ denote, respectively, the $n$-dimensional Planck mass and 
cosmological constant.  

 In this paper, we freely use the notation and conventions of \cite{Wald}.
In particular, following Sec.~10.2 of \cite{Wald}, we use the one-to-one 
correspondence between tensors in $\Sigma$ and tensors in ${\cal M}$ that 
are invariant under projection to the tangent space to $\Sigma$, i.e., 
tensors $T^{a_1 \cdots a_k}{}_{b_1 \cdots b_l}$ such that 
\beq
T^{a_1 \cdots a_k}{}_{b_1 \cdots b_l} = h^{a_1}{}_{c_1} \cdots h^{a_k}{}_{c_k}
h_{b_1}{}^{d_1} \cdots h_{b_l}{}^{d_l} T^{c_1 \cdots c_k}{}_{d_1 \cdots d_l} 
\, .
\eeq

 The third term in (\ref{action}) containing the four-dimensional curvature 
is often missing from the action, or its contribution is missing from the 
equations of motion.  However, in general, this term seems to be essential 
since it is generated as a quantum correction to the matter action in 
(\ref{action}).  Note that this quantum correction typically involves an 
infinite number of terms of higher order in curvature (a similar situation 
in the context of the AdS/CFT correspondence is described in \cite{HHR}).  
Thus, we assume that such terms are present and retain only the lowest-order 
ones in (\ref{action}).

 In this paper, we are interested only in the metric equation, so the 
Lagrangians for the matter fields $\Phi$ and $\phi$ will not be specified.  
The first variation of action (\ref{action}) with respect to the metric 
$g_{ab}$ is equal to\footnote{Those interested in the derivation of 
(\ref{variation}) may look into the appendix.}
\begin{eqnarray} \label{variation}
\delta S &=& M_5^3 \int_{\cal M} 
\left({}^{(5)}G_{ab} + \Lambda_5 g_{ab} \right) \delta g^{ab} 
-  \int_{\cal M} T_{ab} \delta g^{ab}
+ M_5^3 \int_{\Sigma} S_{ab} \delta h^{ab} \nonumber \\
&+& M_4^2 \int_{\Sigma} 
\left({}^{(4)}G_{ab} + \Lambda_4 h_{ab} \right) \delta h^{ab} 
- \int_{\Sigma} \tau_{ab} \delta h^{ab} \, ,
\end{eqnarray}
where ${}^{(n)}G_{ab}$ denotes the $n$-dimensional Einstein's tensor, 
$S_{ab} \equiv K_{ab} - K h_{ab}$, and $T_{ab}$ and $\tau_{ab}$ define, 
respectively, the five-dimensional and four-dimensional stress-energy 
tensors of matter.  Note that the variation $\delta h^{ab}$
on the brane is not an independent quantity, but is 
completely and uniquely determined by $\delta g^{ab}$.  For simplicity, we 
also assume that the Lagrangian $L_5(g_{ab}, \Phi)$ does not contain
derivatives of the metric $g_{ab}$, the presence of which might  
contribute to the surface terms in (\ref{variation}).

 On an extremal field configuration, variation (\ref{variation}) 
is equal to zero for arbitrary variations of the metric $g_{ab}$.  
Considering variations that leave the induced metric on $\Sigma$ intact, 
i.e., for which $h^a{}_c h^b{}_d \delta h^{cd} \equiv 0$ and, hence, the 
surface integrals in (\ref{variation}) vanish, we obtain the 
equation of motion in the five-dimensional bulk:
\beq \label{bulk}
{}^{(5)}G_{ab} + \Lambda_5 g_{ab} = {1 \over M_5^3} T_{ab} \, . 
\eeq
It is important to stress that the Gibbons--Hawking boundary term [the second 
term in the square brackets in action (\ref{action})] is required to obtain 
this equation in a consistent way \cite{GH}.  Now, considering 
arbitrary variations of the metric $g_{ab}$ and taking into account equation
(\ref{bulk}), we obtain the equation of motion on the boundary $\Sigma$ 
in the form
\beq \label{brane}
{}^{(4)}G_{ab} + \Lambda_4 h_{ab} + M S_{ab} = {1 \over M_4^2} \tau_{ab} \, ,
\eeq
where $M = M_5^3 / M_4^2$.  It is the presence of the tensor 
$S_{ab}$ in the equation of motion (\ref{brane}) that makes the dynamics
on the brane $\Sigma$ unusual. 

 One of the Gauss--Codacci relations, namely, 
\beq
D_a S^a{}_b = {}^{(5)}R_{cd} n^d h^c{}_b \, , 
\eeq
where $D_a$ is the (unique) derivative on the brane $\Sigma$ 
associated with the induced metric $h_{ab}$, together with 
equation (\ref{brane}) and the bulk equation (\ref{bulk}) imply the
relation
\beq \label{conserve}
D_a \tau^a{}_b = T_{cd} n^d h^c{}_b \, . 
\eeq
Thus, the four-dimensional stress-energy tensor is covariantly conserved 
if and only if the right-hand side of (\ref{conserve}) is vanishing at 
the brane, in particular, if the stress-energy tensor $T_{ab}$ of the 
five-dimensional matter is a linear combination of $g_{ab}$ and $h_{ab}$ 
at the brane.

 Thus far, we considered $\Sigma$ to be a boundary of a five-dimensional 
manifold ${\cal M}$.  However, the theory can easily be extended to the 
case where $\Sigma$ is embedded into ${\cal M}$.  In this case, it can be 
regarded as a common boundary of two pieces ${\cal M}_1$ and ${\cal M}_2$ 
of ${\cal M}$, and we should simply add the actions of the form of the first 
term in (\ref{action}) for these two pieces.  In varying the resulting action, 
we must respect the condition that the metrics induced on the brane $\Sigma$
by the metrics of these two pieces coincide; however, the extrinsic 
curvatures of $\Sigma$ in ${\cal M}_1$ and in ${\cal M}_2$ are allowed to be 
different.  Equation (\ref{bulk}) remains valid in the bulk, and equation 
(\ref{brane}) will be modified to 
\beq \label{brane1}
{}^{(4)}G_{ab} + \Lambda_4 h_{ab} + M \left(S_{ab}^{(1)} + 
S_{ab}^{(2)}\right) = {1 \over M_4^2} \tau_{ab} \, ,
\eeq
where the tensors $S_{ab}^{(1)}$ and $S_{ab}^{(2)}$ are constructed with the 
use of the respective extrinsic curvatures.  The analog of equation 
(\ref{conserve}) also can easily be derived. 

 If several branes are embedded into the manifold ${\cal M}$, 
equations of the form (\ref{brane1}) are valid for each of these branes.
If a brane is a common boundary of more than two bulk manifolds, the 
corresponding number of similar terms will be present inside the brackets 
of (\ref{brane1}).

 Neglecting the third term in action (\ref{action}) amounts to taking
the limit of $M_4 \to 0$.  In this case, equation (\ref{brane1}) 
reduces to the Israel's junction condition \cite{Israel}
\beq
M_5^3 \left(S_{ab}^{(1)} + S_{ab}^{(2)}\right) = \tau_{ab} \, . 
\eeq

 On the other hand, taking the limit of $M_5 \to 0$ is equivalent to 
setting $M = 0$ in equation (\ref{brane}) or (\ref{brane1}).  In this 
limiting case, the influence of the five-dimensional bulk vanishes, 
and we obtain the standard equation of general relativity.

\section{Cosmological example}

 As an illustration, we consider a particular example of the cosmological 
situation described in \cite{BV}.  We take the five-dimensional metric in
the static spherically-symmetric form
\beq \label{metric}
ds_5^2 = - f(r) dt^2 + dr^2 / f(r) + r^2 d\Omega_{(3)} \, , 
\eeq
where $d\Omega_{(3)}$ is the metric of the unit three-sphere.  In the case 
of vanishing stress-energy tensor of the five-dimensional matter, as a 
solution of (\ref{bulk}) we can take 
\beq \label{f}
f(r) = 1 - \alpha r^2 \, , 
\eeq
where $\alpha = \Lambda_5 / 6$.  The brane $\Sigma$ is taken to be 
spherically-symmetrically embedded into this manifold 
according to the law $r = a(t)$.  We then discard the exterior $r > a(t)$ 
and consider the resulting space with $\Sigma$ as a boundary.  The tensor 
$S_{ab}$ for this boundary can easily be calculated.  Its nonzero components 
have the form (see also \cite{BV})
\beq
S^0{\!}_0 = {} - {3 \sqrt{f(a) + \dot a^2} \over a} \, , \qquad  
S^i{\!}_j = {} - \delta^i{\!}_j {1 \over a^2 \dot a} {d \over d \tau}
\left(a^2 \sqrt{f(a) + \dot a^2}\right), 
\eeq
where $i, j = 1, 2, 3$ label the coordinates on the unit three-sphere
and the overdot denotes the derivative with respect to the cosmological 
time $\tau$ on the brane, in terms of which the induced metric is given by 
the line element
\beq
ds_4^2 = - \left[f(a) - {1 \over f(a)} \left({da \over dt}\right)^2
\right] dt^2 + a^2 d\Omega_{(3)} = - d \tau^2 + a^2 d \Omega_{(3)} \, . 
\eeq

 Equation (\ref{brane}) then yields the following two equations:
\beq \label{time}
{1 + \dot a^2 \over a^2} + {M \sqrt{f(a) + \dot a^2} \over a} = 
\lambda + \kappa \rho \, , 
\eeq \beq \label{space}
{2 \ddot a \over a} + {1 + \dot a^2 \over a^2} 
+ {M \over a^2 \dot a} {d \over d \tau}
\left(a^2 \sqrt{f(a) + \dot a^2}\right) = 3 \lambda - 3 \kappa p \, ,  
\eeq
where we made the notation $\lambda = \Lambda_4 / 3$, 
$\kappa = 1 / 3 M_4^2$, and $\rho$ and $p$ denote the standard components
of the four-dimensional stress-energy tensor (that may include its own 
cosmological-constant contribution).  Introducing the Hubble 
parameter $H \equiv \dot a / a$, we have from (\ref{time})
\beq \label{hubble}
H^2 + {f(a) \over a^2} = \left[\left({M^2 \over 4} + {f(a) - 1 \over a^2} + 
\lambda + \kappa \rho\right)^{1/2} - {M \over 2} \right]^2  . 
\eeq
Substituting this into equation (\ref{space}), we obtain the conservation law
\beq \label{conserve1}
\dot \rho + 3 H (\rho + p) = 0 \, , 
\eeq
in accordance with the general equation (\ref{conserve}). 
Together with the equation of state (e.g., in the form $p = w \rho$), 
equations (\ref{hubble}), (\ref{conserve1}) constitute a closed system of 
cosmological equations.

 If we neglect the third term in action (\ref{action}) and thus take the 
limit of $M_4 \to 0$, equation (\ref{hubble}) will become 
\beq \label{nonstandard}
H^2 + {f(a) \over a^2} = \left({\rho \over 3 M_5^3} \right)^2 , 
\eeq
with the quadratic dependence of the right-hand side on the energy density, 
noted previously \cite{LOW,BDL,BDEL,BV}.  This result is approximately valid 
for finite values of $M_4$ provided 
$\left|(f(a) - 1) / a^2 + \lambda\right| \ll \kappa \rho \ll M^2$.  

 In the limit of $M_5 \to 0$, which corresponds to 
$M \to 0$, the influence of the five-dimensional bulk vanishes, 
and we recover the equation of the standard cosmology
\beq \label{standard}
H^2 + {1 \over a^2} = \lambda + \kappa \rho \, .
\eeq
This equation is approximately valid for finite values of $M_5$ provided 
$M^2 \ll (f(a) - 1) / a^2 + \lambda + \kappa \rho$.  

 We see that, in principle, the standard regime (\ref{standard}) might be 
realised at the early stages of the universe expansion, the standard 
theory of nucleosynthesis thus remaining intact, while, at later stages, the 
universe might evolve according to the nonstandard law (\ref{nonstandard}). 

 The system of equations (\ref{time}),
(\ref{space}) with the function $f(r)$ given by (\ref{f}) admits a solution 
of the static empty (that is, $\rho = p = 0$) universe with the scale factor
\beq \label{scale}
a_0 = \left( \alpha + {M^2 \over 4} \right)^{- 1/2} \, , 
\eeq
provided the value of $M_5$ (hence, also of $M$) is negative and the values 
of parameters are tuned so that
\beq \label{tune}
\lambda = \alpha - {M^2 \over 4} \, . 
\eeq
Note that, in the case of negative $\Lambda_5$, solution (\ref{metric}),
(\ref{f}) describes the anti-de~Sitter five-dimensional space.  In this case, 
the value of $\alpha = \Lambda_5 / 6$ is negative, so that the size of the 
four-dimensional static universe given by (\ref{scale}) may be arbitrarily 
large, because the value of $\alpha + M^2 / 4$ may be arbitrarily close 
to zero.  
This is another instance of fine-tuning, similar to that discussed in 
\cite{RS}, that countervails the effect of the four-dimensional 
cosmological constant which, according to (\ref{tune}), must be negative 
in the present case.

 It should be noted, however, that, under relation (\ref{tune}), the system 
of equations (\ref{time}), (\ref{space}) is degenerate at the point 
\beq \label{Cauchy}
a = a_0\, , \qquad \dot a = 0 \, ,
\eeq
in the sense that the second-order derivative term in (\ref{space}) 
vanishes at this point.  This circumstance, 
in particular, has a consequence that the Cauchy problem (\ref{Cauchy})
for system (\ref{time}), (\ref{space}) is ill-defined since, besides the
static solution $a \equiv a_0$, it also has the solution
\beq
a(\tau) = a_0 \cosh \left({\tau \over a_0}\right) \, 
\eeq
describing the de~Sitter spacetime.

 Gravitational perturbations of solutions to theory (1) and the resulting 
effective four-dimensional theory of gravity remain to be investigated.  
In this respect, some important results in the context of the AdS/CFT 
correspondence are discussed in \cite{HHR}.

\section*{Note added}

 After this paper was already submitted to the LANL archive, the author 
became aware of the paper \cite{CH}, in which all the essential results 
of the present work have been obtained.  

\section*{Acknowledgments}

This work was supported in part by the Foundation of Fundamental Research 
of the Ministry of Science of Ukraine under grant No.~2.5.1/003. 

\newpage
\appendix
\section*{Variation of the action for gravity}

 Here, we derive the expression for the first variation of the action
for gravity
\beq \label{a-action}
S_g =  \int_{\cal M} R + 2 \int_{\Sigma} K \, .  
\eeq
Equations of this section will be valid for arbitrary dimension of spacetime.
In contrast with the standard derivation, here we do not assume that the 
variation of $g_{ab}$ vanishes at the boundary $\Sigma$, which is taken to 
be timelike. 

 We start from the standard expression (see, e.g., Appendix~E of \cite{Wald})
\beq \label{var1}
\delta \left( \int_{\cal M} R \right) = \int_{\cal M} 
G_{ab}\, \delta g^{ab} + \int_{\cal M} \nabla^a v_a \, , 
\eeq
where
\beq
v_a = \nabla^b \left(\delta g_{ab} \right) - g^{cd} \nabla_a 
\left(\delta g_{cd} \right) \, .
\eeq
The second integral in (\ref{var1}) can be transformed with the use of 
the Stokes theorem as
\beq \label{a-vn}
\int_{\cal M} \nabla^a v_a = \int_\Sigma v_a n^a \, , 
\eeq
where
\beq 
v_a n^a = n^a g^{bc} \left[ \nabla_c \left( \delta g_{ab} \right) -
\nabla_a \left( \delta g_{bc} \right) \right] = 
n^a h^{bc} \left[ \nabla_c \left( \delta g_{ab} \right) -
\nabla_a \left( \delta g_{bc} \right) \right] \, . 
\eeq

 Then we have
\beq \label{a-deltaK}
\delta K = \delta \left( h^a{}_b \nabla_a n^b \right) = 
\delta h^a{}_b \nabla_a n^b + h^a{}_b (\delta C)^b{}_{ac} n^c
+ h^a{}_b \nabla_a \delta n^b \, , 
\eeq
where 
\beq
(\delta C)^b{}_{ac} = {1 \over 2}\, g^{bd} \left[ \nabla_a 
\left(\delta g_{cd}\right) + \nabla_c \left(\delta g_{ad}\right) - 
\nabla_d \left(\delta g_{ac}\right) \right] \, . 
\eeq
The first term in the right-hand side of (\ref{a-deltaK}) is identically 
zero.  Indeed, we have $\delta n_a = {}- n_a n_b \delta n^b$, so that
\begin{eqnarray}
&{}&\delta h^a{}_b \nabla_a n^b = {}- \left( \delta n^a n_b + n^a \delta n_b 
\right) \nabla_a n^b = {}- \left( \delta n^a - n^a n_c \delta n^c 
\right) n_b \nabla_a n^b \nonumber \\ &{}&= {}- \delta n^c h^a{}_c n_b 
\nabla_a n^b = {}- \delta n^c n_b K_c{}^b = 0 \, . 
\end{eqnarray}

 Thus, variation of the second term of (\ref{a-action}) is 
\beq \label{a-varK}
\delta \left( 2 \int_\Sigma K \right) = \int_\Sigma \left[ n^c h^{ab} 
\nabla_c \left( \delta g_{ab} \right) + 2 h^a{}_b \nabla_a \delta n^b 
- K h_{ab} \delta h^{ab} \right] \, ,  
\eeq
where the last term in the square brackets stems from the variation of the 
volume element $\sqrt{- h}\, d^4 x$ in the integral over $\Sigma$.  

 The total boundary term in the variation of action (\ref{a-action})  
is given by the sum of (\ref{a-vn}) and (\ref{a-varK}) with the result
\beq
(\mbox{Boundary term}) = 
\int_\Sigma \left[ n^a h^{bc} \nabla_c \left( \delta g_{ab} \right) + 
2 h^a{}_b \nabla_a \delta n^b - K h_{ab} \delta h^{ab} \right] \, .
\eeq

 We transform the first term in the integrand of the last expression:
\beq
n^a h^{bc} \nabla_c \left( \delta g_{ab} \right) = h^{bc} \nabla_c \left(
n^a \delta g_{ab} \right) - h^{bc} \nabla_c n^a \delta g_{ab} = h^{bc} 
\nabla_c \left( n^a \delta g_{ab} \right) + K_{ab} \delta h^{ab} \, .
\eeq
Then
\beq \label{a-boundary}
(\mbox{Boundary term}) = \int_\Sigma \left[
h^{bc} \nabla_c \left( n^a \delta g_{ab} \right) + 
2 h^a{}_b \nabla_a \delta n^b \right] + \int_\Sigma  
\left(K_{ab} - K h_{ab}\right) \delta h^{ab} \, .
\eeq

 Now we show that the integrand of the first integral in (\ref{a-boundary}) 
is a divergence, so that this integral vanishes for 
variations of $g_{ab}$ with compact support in $\Sigma$.  Indeed,
\begin{eqnarray} \label{a-divergence}
&{}&h^{bc} \nabla_c \left( n^a \delta g_{ab} \right) + 2 h^a{}_b \nabla_a 
\delta n^b = h^{bc} \nabla_c \left( \delta n_b - g_{ab} \delta n^a \right)
+ 2 h^a{}_b \nabla_a \delta n^b \nonumber \\ 
&{}&= h^a{}_b \nabla_a \left( g^{bc} \delta n_c + \delta n^b \right) 
= h^a{}_b \nabla_a \left(h^b{}_c \delta n^c \right) 
= D_b \left(h^b{}_c \delta n^c \right) \, , 
\end{eqnarray}
where $D_a$ is the (unique) derivative on $\Sigma$ associated with the 
induced metric $h_{ab}$, and the last equality in (\ref{a-divergence}) 
is valid by virtue of Lemma~10.2.1 of \cite{Wald}.

 As a final result, we obtain 
\beq
\delta S_g = \int_{\cal M} G_{ab}\, \delta g^{ab} + \int_\Sigma  
\left(K_{ab} - K h_{ab}\right) \delta h^{ab} \, .
\eeq


\begin{thebibliography}{99}

\bibitem{vdBDBL}
C.~van~de~Bruck, M.~Dorca, R.~H.~Brandenberger, and A.~Lukas, 
{\it Cosmological Perturbations in Brane-World Theories: Formalism}, 
hep-th/0005032.

\bibitem{LOW}
A.~Lukas, B.~A.~Ovrut, and D.~Waldram, hep-th/9902071, Phys.\@ Rev.\@ D 
{\bf 61}, 023506 (2000).

\bibitem{BDL}
P.~Bin\'etruy, C.~Deffayet, and D.~Langlois, hep-th/9905012, Nucl.\@ Phys.\@ 
B {\bf 565}, 269 (2000).

\bibitem{BDEL}
P.~Bin\'etruy, C.~Deffayet, U.~Ellwanger, and D.~Langlois, hep-th/9910219, 
Phys.\@ Lett.\@ B {\bf 477}, 285 (2000).

\bibitem{BV}
C.~Barcel\'o and M.~Visser, {\it Living on the Edge: Cosmology on the 
Boundary of Anti-de~Sitter space}, hep-th/0004056.

\bibitem{CGKT}
C.~Cs\'aki, M.~Graesser, C.~Kolda, and J.~Terning, hep-ph/9906513, 
Phys.\@ Lett.\@ B {\bf 462}, 34 (1999).

\bibitem{CGS}
J.~M.~Cline, C.~Grojean, and G. Servant, hep-ph/9906523, Phys.\@ Rev.\@ 
Lett.\@ {\bf 83}, 4245 (1999). 

\bibitem{CGRT}
C.~Cs\'aki, M.~Graesser, L.~Randall, and J.~Terning, {\it Cosmology
of Brane Models With Radion Stabilization}, hep-ph/9911406.

\bibitem{Wald}
R.~M.~Wald, {\it General Relativity}, The University of Chicago Press, 
Chicago (1984).

\bibitem{HHR}
S.~W.~Hawking, T.~Hertog, and H.~S.~Reall, {\it Brane New World}, 
hep-th/0003052.

\bibitem{GH}
G.~W.~Gibbons and S.~W.~Hawking, Phys.\@ Rev.\@ D {\bf 15}, 2752 (1977).

\bibitem{Israel}
W.~Israel, Nuovo Cimento B {\bf 44}, 1 (1966); Errata---{\it ibid.} 
{\bf 48}, 463 (1967).

\bibitem{RS}
L.~Randall and R.~Sundrum, hep-ph/9905221, Phys.\@ Rev.\@ Lett.\@ {\bf 83},
3370 (1999); \mbox{hep-th/9906064}, Phys.\@ Rev.\@ Lett.\@ {\bf 83}, 4690 (1999). 

\bibitem{CH}
H.~Collins and B.~Holdom, {\it Brane Cosmologies Without Orbifolds}, 
hep-ph/0003173.

\end{thebibliography}
\end{document}